\documentclass[final,5p,times]{elsarticle}

\usepackage{graphicx}
\usepackage{graphics}
\usepackage{amssymb}

\def\apj{{\it Astrophys.~J.}}

\def\prd{{\it Phys.~Rev.~D.}}
\def\prl{{\it Phys.~Rev.~Lett.}}
\def\plb{{\it Phys.~Letts.~B}}
\def\mnras{{\it Mon.~Not. Roy.~Astr.~Soc.}}
\def\ijmp{{\it Int.~J.~ Mod. Phys.}}

\begin{document}

\begin{frontmatter}

\title{Constraints on Dark Energy and Modified
Gravity models by the Cosmological Redshift Drift test}
\author[dpj]{Deepak Jain}
\ead{djain@ddu.du.ac.in}

\author[skj]{Sanjay Jhingan\corref{cor1}}
\ead{sanjay.ctp@jmi.ac.in}

\address[dpj]{ Deen Dayal Upadhyaya College, University of Delhi, New
Delhi 110015, India}

\address[skj]{Centre for Theoretical Physics, Jamia Millia Islamia, New
Delhi 110025, India}

\cortext[cor1]{Sanjay Jhingan}

\begin{abstract}
We study cosmological constraints on the various accelerating models
of the universe using the time evolution of the cosmological
redshift of distant sources. The important characteristic of this
test is that it directly probes the expansion history of the
universe. In this work we analyze the various models of the universe
which can explain the late time acceleration, within the framework
of General Theory of Relativity (GR) (XCDM, scalar field
potentials) and beyond GR ($f(R)$ gravity model).
\end{abstract}

\end{frontmatter}

\section{Introduction}

The recent accelerated expansion of the universe is one of the most
important discovery in the cosmology. Whether this observed
acceleration is due to some new hypothetical energy component with
large negative pressure (dark energy) within the framework of
General  Theory of Relativity, or due to modification in the GR at
the cosmological distances (modified gravity), is not known.
Therefore many models have been proposed in literature to understand
the origin and nature of this present acceleration~\cite{da}.

Basically the cosmic acceleration affects the expansion history of
the universe. Therefore to understand the true nature of this
driving force, mapping of the cosmic expansion of the universe is
very crucial~\cite{li}. Hence, we require various observational
probes in different redshift ranges to understand the expansion
history of the universe. The observational tools for probing the
cosmic acceleration broadly fall into two categories: Geometrical
and Dynamical probes. A {\it Geometrical probe} deals with large
scale distances and volume which include luminosity distance
measurements of SNe Ia, angular diameter distance from first CMB
acoustic peak, Baryon Acoustic Oscillations (BAO) etc.. A {\it
Dynamical probe} investigates the growth of matter density
perturbations that give rise to the  large scale structure such as
galaxies, clusters of galaxies etc. in the universe.

Using supernovae as standard candles is a popular method of
constraining the properties of dark energy. Though this method is
very simple and useful in constraining the various dark energy
models, at present the luminosity distance measurements suffers from
many systematical uncertainties like extinction by dust,
gravitational lensing etc.~\cite{nordin}. On the other hand,
measuring the expansion history from growth of matter perturbations
also has its limitations. It requires prior information of exact
value of matter density, initial conditions, cosmological model
etc.~\cite{li,l}. So the question arises, ``Is there any probe which
is simple, depends on fewer priors and assumptions?'' The possible
probe is ``Cosmological Redshift Drift'' (CRD) test which maps the
expansion history of the universe directly.

The CRD test is based on very simple and straightforward physics.
However, observationally it is a very challenging task and requires
technological breakthroughs~\cite{odorico}. The most remarkable
feature of this probe is that it measures the dynamics of the
universe directly - the Hubble expansion factor. This property makes
it very special and unique. Further, it assumes that the universe is
homogeneous and isotropic at the cosmological scales [for more
details see ref.~\cite{lis}]. The time drift of the cosmological
redshift probes the universe in the redshift $2 < z < 5$, whereas
the other cosmological tests based on SNe Ia, BAO, weak lensing,
number counts of clusters etc. have not penetrated beyond $z = 2$.
The other advantage of this tool is that it has controlled
systematical uncertainties and evolutionary effects of the sources.

The aim of this letter is to employ CRD test to constrain various
accelerating models both within the framework of GR and beyond GR.
We have used \emph{simulated data points} for redshift drift
experiment generated by Monte Carlo simulations with the assumption
of standard cosmological model ($\Lambda$ CDM) as
reference~\cite{lis,joe}. We put constraints on the Quintessence
models (based on scalar field potentials) like PNGB, inverse  power
law and exponential potentials. The dark energy parametrization
which has a variable equation of state is also investigated. We also
put a bound on the $f(R)$ gravity models i.e., Starobinsky model
using the Cosmological Redshift Drift test.

The letter is organized as follows. In Section 2 we review the
theoretical basis of Cosmological Redshift Drift test. We also
present here methodology and data used for this work. Various models
which can explain late time accelerated expansion of the universe
are described in Section 3. Last section contains a summary and
discussion of results.

\section{\bf Cosmological Redshift Drift Test }

\subsection{Theory}

A test which can trace the dynamical expansion history of the
universe was proposed by Sandage ~\cite{san}. The expansion of the
universe is expressed in terms of a scale factor, $a(t)$. Therefore,
the time evolution of the scale factor, or change in redshift,
${\dot z}$, directly measures the expansion rate of the universe.
The redshift, $z$, of an object as determined today will be
different from its measured value after a time interval of several
years. Sandage also stressed on the fact that the redshift drift
signal, ${\dot z}$, is very small. The significance of this tool has
been discussed by several authors ~\cite{ot}. Loeb was the first to
suggest the possibility of measuring the redshift drift by observing
Ly$\alpha$ absorption lines in the spectra of quasars
(QSOs)~\cite{lo}. This reinforced the importance of this probe.

The observed redshift of a distant source is given by
\begin{equation}
z(t_0)\, = \, \frac{a(t_0)}{a(t_s)}- 1
\end{equation}
where $t_s$ is the time at which the source emitted the radiation
and $t_0$ is the time at which the observation is made. In writing
the above expression, we ignore any peculiar motion of the object.
The redshift of the source after the time interval of $\Delta t_0$
becomes
\begin{equation}
z(t_0 +\Delta t_0)= \frac{a(t_0 +\Delta t_0)}{a(t_s + \Delta t_s)}-
1
\end{equation}
where $\Delta t_s$ is the emission time interval for the source. In
the first order approximation we can write
\begin{equation}
\frac{\Delta z}{\Delta t_0} \,\approx \, \frac{ ({\dot a(t_0)}-\dot
a(t_s) )}{a(t_s)}
\end{equation}
or
\begin{equation}
\dot{z} = H_0\left[ 1 + z- \frac{H(z)}{H_0}\right]
\end{equation}
The above equation is also known as McVittie equation~\cite{mc}.
This clearly shows that ${\dot z}$ traces $H(z)$, which is the
Hubble factor at redshift $z$. As stated earlier ${\dot z}$ measures
the rate of expansion of the universe: ${\dot z} > 0$ and $ < 0$
indicates the accelerated and decelerated expansion of the universe,
respectively. For a coasting universe  ${\dot z} = 0$. The redshift
variation is related to the apparent velocity shift of the source:
\begin{equation}
\Delta {\mathrm v} = c\frac{\Delta z}{(1+z)} .
\end{equation}
Thus we can write
\begin{equation}
{\dot {\mathrm v}} = \frac{c H_0}{(1+z)} \left[ 1+z
-\frac{H(z)}{H_0} \right]
\end{equation}
where ${\dot {\mathrm v}}= \Delta {\mathrm v}/\Delta t_0$ and $H_0 =
100 \, h\,\, km/s/Mpc$. In a standard cosmological model
($\Lambda$CDM), with a time interval of $ \Delta t_0 = 10$ yr, the
change in redshift is $\Delta z \approx 10^{-9}$, for a source at
redshift $ z = 4$. The corresponding shift in the velocity is of the
order of $\Delta{\mathrm v} \approx 6 $ cm/s. To measure this weak
signal, Loeb pointed out that observation of the Ly$\alpha$ forest
in the QSO spectrum for a decade might allow the detection of signal
of such a tiny magnitude.

\subsection {Data}

In the near future, a new generation Extremely Large Telescope (ELT,
25 - 42 m diameter) equipped with a high resolution, extremely
stable and ultra high precision spectrograph (CODEX) should be able
to measure such a  small cosmic signal. The CODEX (COsmic Dynamics
EXperiment) operates in the spectral range of 400-680 nm with
resolving power R = 150000. Several groups have performed Monte
Carlo (MC) simulations of quasars absorption spectra~\cite{mcs,pa}
and obtained the $\dot z$ measurements. In this work we have used
three sets of data (8 points) for redshift drift experiments
generated by MC simulations~\cite{lis,joe}.

These three datasets are generated by three different approaches
\cite{lis}. The data set with error bars are generated by assuming
the total duration of 20 years for observations and  standard input
cosmological model with $H_0 \, = \, 70$\, km/s/Mpc, $\Omega_m =
0.3$ and $ \Omega_{\Lambda} = 0.7$. In every approach it is assumed
that normalized observational set-up parameter, $O$, equal to 2.
This parameter controls  the telescope size, efficiency and
integration time. In the {\it first} approach, the data points are
selected by smallest value of $\sigma_{\mathrm v}$. In this approach
set of 20 QSOs are distributed in the four equally sized redshift
bin.  The {\it second} set of data points are generated by selecting
the targets by larger value of $|{\dot{\mathrm
v}}|/\sigma_{{\dot{\mathrm v}}}$. In this case,   $N_{QSOs} = 10$
are distributed in two redshift bins. Finally the {\it last} two
data points  are generated by increasing the sensitivity of data
towards $\Omega_{\Lambda}$.

In principle, redshift experiment involve the measurement of
velocity shift between pair of spectra of same object separated by
large time interval of many years. As mentioned earlier, There are
many candidates available for this measurement but the most
promising one is the absorption features of the Ly$\alpha$ forest in
the high redshift QSOs. The main reason is that the peculiar motion
associated with intervening gas is negligible as compare to cosmic
signal and numerous lines are present in the single spectrum. Using
MC simulations, the desired statistical error on velocity shift can
be written as
$$\sigma_{\mathrm v} \,=\, 1.35 \left(\frac{S/N}{2370}\right )^{-1}\,
\left(\frac{N_{QSOs}}{30}\right)^{-1/2}\,
\left(\frac{1+z_{QSOs}}{5}\right)^{-a}\, cm/s
$$
with $a = 1.7$ for $ z\le 4$ and $a = 0.9$ for $ z > 4$, where $S/N$
is the signal-to-noise ratio is of the order of 13500 per 0.0125
${{A}^{\circ}}$ pixel, $N_{QSOs}$ is the total number of quasar
spectra observed and $z_{QSOs}$ is their redshift. QSOs are the
brightest sources which exist even at the redshift close to 6 and
since we are observing it from ground, the existence of atmosphere
put the lower limit of redshift $z \gtrsim 2$. Using full MC
simulation to achieve a radial velocity accuracy of 3 cm/s, can be
obtained  with 3200 hours of observation with 42 m ELT. For more
details see Ref. \cite{lis}.

\subsection{Method}
We perform the $\chi^2$ analysis to find the best fit values of the
cosmological parameters and to find the bounds on them,
\begin{equation}
\chi^2(p) = \sum_{k = 1}^{8} \frac{({ \dot{\mathrm v}_{th}}(z_k,h,
p) - {\dot{\mathrm v}}_{MC}(z_k))^2} { {\sigma_{k}(z_k)}^2 }
\end{equation}
where $p$ are the model parameters. Here $\dot{\mathrm v}_{th}$ and
${\dot{\mathrm v}}_{MC}$ are the expected and the simulated values
of the velocity drift respectively. The error bars on the velocity
drift are denoted by $\sigma_{k}$. To draw the likelihood contours
at 1, 2 and 3$\sigma$, $\Delta {\chi}^2 = {\chi}^2- {\chi}^2_{min} =
$ 2.30, 6.17, 11.8 respectively in two-dimensional parametric space.

\section{\bf  Models}
In this work, we investigate various models of the universe which
can explain late time acceleration both within the framework of GR
(Einstein gravity) and beyond GR (modified gravity).

\subsection{Models Based on Einstein Gravity}

\subsubsection{XCDM Model} One of the widely studied model of dark energy
is XCDM parametrization. In this model the dark energy is
characterized by an equation of state $w_x = p/{\rho}$, where $p$
and $\rho$ are  pressure and energy density respectively. Further,
the equation of state does not evolve with time. For acceleration in
this model $w_x < -1/3$. The Friedmann equation in this model is:
\begin{equation}
{\left [\frac{H(z)}{ H_0 }\right ]^2}\, = \, \Omega_m\,(1 +z)^3 \, +
\, \Omega_x(1+ z)^{3(1 + w_x)}
\end{equation}
Here we assume the universe is flat i.e., $\Omega_m + \Omega_x = 1$.
$\Omega_m$ and $\Omega_x$ are the fractional matter and dark energy
densities at the present epoch respectively.

\subsubsection{Scalar Field Cosmological Models}
In order to get Hubble parameter, $H(z)$, in the scalar field
cosmology, we have to solve the following equations of motion. The
Einstein field equations can be expressed as:
\begin{equation}
{\dot{H}} =-\frac{3}{2}\,H^2 - 2\pi G\, {{\dot{\phi}}}^2\,
+\, 4\pi G\,V(\phi)
\end{equation}
Here dots are derivative w.r.t. time. The scalar field equation of
motion is
\begin{equation}
\ddot{\phi}\, = \, -3H\,{{\dot{\phi}}}\,-\,{V'{(\phi)}}
\end{equation}
where prime stands for a derivative w.r.t. the scalar field $\phi$,
and the Hubble parameter is
\begin{equation}
H^2 \, =\, \frac{8\pi\,G}{ 3}\left[ \rho_m \, + \, \frac{{\dot
{\phi}}^2}{2} \,+\, V(\phi) \right] .
\end{equation}
Here $\rho_m$ is energy density of matter and $m_{pl} = G^{-1/2}$ is
the Planck mass. The equation of state $\omega$ is defined as
\begin{equation}\label{eqn_state}
\omega = \frac{{\dot \phi}^2 - 2 V(\phi)}{{\dot \phi}^2 + 2 V(\phi)}
.
\end{equation}
We analyse three scalar field potential models.

\begin{itemize}
\item  {\underline{Inverse power law potential}}: In this model
the scalar field potential is of the form:
\begin{equation} V(\phi)\, = \,
\frac{k}{ {32\pi\,G^2}}\, \left(\frac{1}{{\sqrt{16\pi\,
G}}\,\phi}\right)^{\gamma}
\end{equation}
where both $k$ and $\gamma$ are positive and dimensionless
constants.  $G$ is the gravitational constant~\cite{ra}. This
potential shows the tracking behaviour in which scalar field start
from wide range of initial set of conditions and in the late time it
approaches the cosmological constant. The
Friedmann equation in terms of dimensionless parameters becomes [for
details see ref.~\cite{waga}]
\begin{equation}
Y^2 = \left(\frac{H}{ H_0}\right)^2=\frac{X^2 }{ 12}\, +\,
\frac{{k\, m_{pl}^2} }{ {12 \,H_0^2 }}\, \, Z^{-\gamma} \, + \,
\Omega_m
\end{equation}
where $ {k} m_{pl}^2/ {H_0^2} = 36 {K}/h^2 $, with $K
>0$. The dimensionless variables are defined as:
\[X = \frac{4 \sqrt{\pi}}{H_0m_{pl}}{\dot \phi} ,\,  Y = \frac{H}{
H_0}, \, Z = \frac{4{\sqrt{\pi}}\phi}{m_{pl}} \quad .
\]
With these definitions we can write the equations of motion in the
following first order form
\begin{eqnarray}\label{sys_power_law}
\frac{dX}{dz} &=& \frac{1}{(1+z)} \left[3 X - \frac{\gamma k
m_{pl}^2}{2 H_0^2} \frac{Z^{-(\gamma + 1)}}{Y} \right] \nonumber
\\
\frac{dY}{dz} & = & \frac{1}{(1+z)} \left[\frac{3}{2} Y +\frac{1}{8}
\frac{X^2}{Y} -\frac{k m_{pl}^2}{8 H_0^2} \frac{Z^{-\gamma}}{Y}\right] \\
\frac{dZ}{dz} & = & -\frac{1}{(1+z)} \left[\frac{X}{Y}\right]
\nonumber
\end{eqnarray}
We solve the above system for parameter values $k = K \times
10^{-120},  \quad H_0 = 1.67 \times 10^{-33} h \,eV, \quad m_{pl} =
1.2 \times 10^{28} eV$, and $h =0.7$. For these values for given set
of parameters we choose initial values (redshift , $z \sim 3000$)
$Z_i = 2.1, X_i =10^{-9}$. Now for different values of the parameter
$K$, we generate initial data sets by choosing $Y_i$ in such a way
that $Y=1$ at $z=0$.

The variation of velocity drift w.r.t. source redshift is shown in
Fig. \ref{power-law}.

\begin{figure}[ht]
%\framebox{
\includegraphics[width=5.80cm,angle=-90]{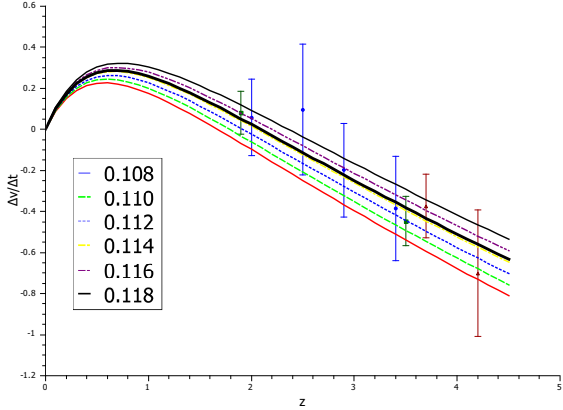}
%}
\caption{Velocity drift (cm/sec/yr) versus redshift for inverse power law potential with
different values of $K$ ($\gamma$ is fixed at 0.001). The bold black curve correspond to the $ \Lambda CDM$ model.
}\label{power-law}
\end{figure}

\item {\underline{Exponential potential}}:  This potential has the
following form
\begin{equation}
V(\phi)\, = \, V_0\,exp(-\phi/f)
\end{equation}
where $V_0$ and $f$ are positive constants~\cite{waga,e}. The
 exponential potential has the capability of producing scaling solutions
 which further scales the background energy density. But it require
 fine tuning of parameters to produce the late time acceleration.
 The
corresponding Hubble parameter in this model is
\begin{equation}
Y^2 = \frac{X^2}{ {6\beta}} + \frac{8\pi }{ 3}e^{-Z} + \Omega_m
\end{equation}
where $ \beta = m_{pl}^2/( 8 \pi f^2)$. The dimensionless variables
are defined as:

\[X = \frac{1}{{H_0 f}} {\dot \phi}, \; Y = \frac{H}{ H_0}, \;
Z = \frac{\phi}{f} - \ln \left[\frac{V_0 }{{m_{pl}^2H_0^2}}\right].
\]
Field equations can now be written as
\begin{eqnarray}
\frac{dX}{dz} &=& \frac{1}{(1+z)} \left[3X - 8\pi\beta
\frac{e^{-Z}}{Y} \right] \nonumber \\
\frac{dY}{dz} &=& \frac{1}{(1+z)} \left[\frac{3}{2} Y +
\frac{1}{4\beta} \frac{X^2}{Y} - 4\pi \frac{e^{-Z}}{Y} \right]
\\\nonumber \frac{dZ}{dz} &=& -\frac{1}{(1+z)}
\left[\frac{X}{Y}\right]
\end{eqnarray}

We solve the above system for parameter values $H_0 = 1.67 \times
10^{-33} h$  $eV$, $ m_{pl} = 1.2 \times 10^{28} eV$ and $h =0.7$.
For these values for given set of parameters we choose initial
values (redshift , $z \sim 3000$) $Z_i = 1.5, X_i =10^{-9}$. Now
for different values of the parameter $\beta$, we generate initial
data sets by choosing $Y_i$ in such a way that $Y=1$ at $z=0$.

The variation of the velocity drift for exponential potential model
are shown in Fig. \ref{fig:exp-pot}. It is clear from the figure
that after certain value of $\beta$, the universe remains in the
decelerated phase.

\begin{figure}[ht]
\includegraphics[width=5.50 cm,angle=-90]{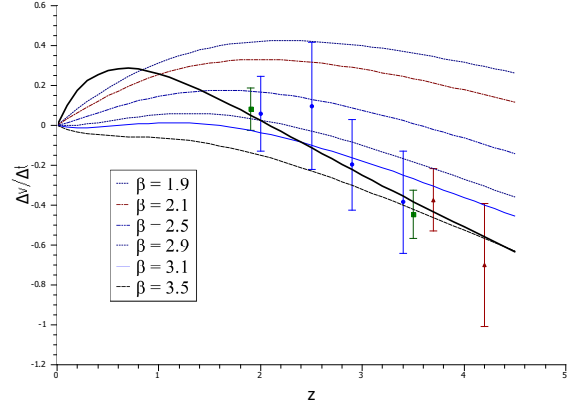}
\caption{Variation of velocity drift (cm/sec/yr) with redshift $z$ for $V=
V_0exp(-\phi/f)$ .  The bold black curve correspond to the $ \Lambda CDM$ model.}\label{fig:exp-pot}
\end{figure}

\item {\underline{PNGB Model}}:  In this (Pseudo-Nambu-Goldstone Bosons) model
the scalar field potential has the following functional form
\begin{equation}
V(\phi)\, = \, M^4\, \left(1+ \, cos({\phi/ f})\right)
\end{equation}
where $M$ and $f$ are positive constants~\cite{waga,ng,pngb}. The main
motivation to study this potential  is because of its special
properties which not only explain late time acceleration but its
ability to offer solution to the cosmic coincidence problem.
The Hubble parameter for PNGB model in the term of dimensionless
parameters is
\begin{equation}
Y^2 \, = \, \frac{4\pi}{3}\alpha^2\, X^2 \, + \,1 + \cos(Z) +
\Omega_m
\end{equation}
where
\begin{equation}
X = \frac{\dot \phi}{H_0 f}, \; Y = \frac{H}{H_0}, \; Z= \frac{\phi
}{f}
\end{equation}
and $ \alpha = f/m_{pl}$. The full dynamical system in terms of
variables $X, Y$ and $Z$ can be written analogues to earlier two
cases as
\begin{eqnarray}
\frac{dX}{dz} &=& \frac{1}{(1+z)}\left[ 3X - \frac{m^4}{\alpha^2}
\left(\frac{\sin Z}{Y} \right) \right] \nonumber\\
\nonumber\frac{dY}{dz} &=& \frac{1}{(1+z)} \left[\frac{3}{2} Y
+{2\pi \alpha^2}
\frac{X^2}{Y} -{4\pi m^4}\left(\frac{1+\cos Z}{Y} \right) \right] \\
\frac{dZ}{dz} &=& -\frac{1}{(1+z)}\left[\frac{X}{Y} \right] .
\end{eqnarray}
where $m^4 = M^4/(m_{pl} H_0)^2 $. We choose initial values
(redshift , $z \sim 3000$) $Z_i = 1.5, X_i = 10^{-9}$. Now for
different values of the parameter $\alpha$, we generate initial data
sets by choosing $Y_i$ in such a way that $Y=1$ at $z=0$.

The variation of the velocity drift for PNGB model are shown in Fig.
\ref{PNGB}. It is clear from the figure that after a certain
critical value of $\alpha$, universe always remain in the
deceleration mode when the light is traveling toward us. For smaller
values of $\alpha$, it shows the universe has undergone the
decelerating phase twice.

\begin{figure}[ht]
%\framebox{
\includegraphics[width=5.50cm,angle=-90]{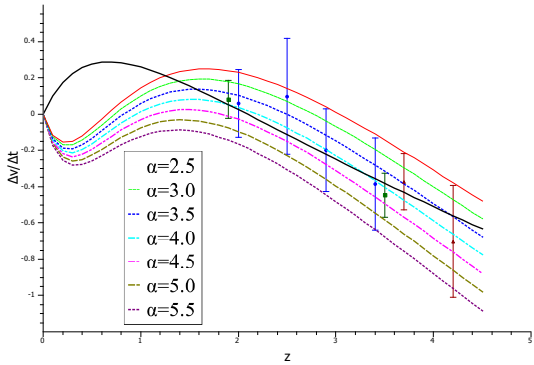}
%}
\caption{Variation of velocity drift (cm/sec/yr) with redshift $z$ for $V= M^4
(1+\cos(\phi/f))$ for  $M= 0.004 {\sqrt{h}} \,eV$. The black curve correspond to the $\Lambda CDM$ model.}\label{PNGB}
\end{figure}

\end{itemize}

\subsection{Modified Gravity Model}

An important class of models which has attracted considerable
attention in past few years is the one that modifies
Einstein-Hilbert action by replacing Ricci tensor by an arbitrary
function of curvature
\begin{eqnarray}
S=\int\left( \frac{f(R)}{16 \pi G} + {\cal L}_m\right)\sqrt{-g}\,
d^4x .
\end{eqnarray}
In this letter we study the $f(R)$ theory model given recently by
Starobinsky, Hu and Sawicki~\cite{s,hu} (see also, \cite{frolov})
\begin{equation}
f(R) \, = \,R \, + \, \lambda R_0\left[\left(1 \, +\, \frac{R^2}{
R_0^2} \right)^{-n} \, \, - \,1 \right]
\end{equation}
where $\lambda$, $n$ and $R_0$ are positive parameters. For
necessary steps leading to calculation of Hubble parameter see Dev
et al. \cite{dev}. These models can evade local gravity constraints and have the capability being distinguished from the cosmological constant.

\begin{figure}[ht]
\centering
%\framebox{
\includegraphics[width=5.5cm,angle=-90]{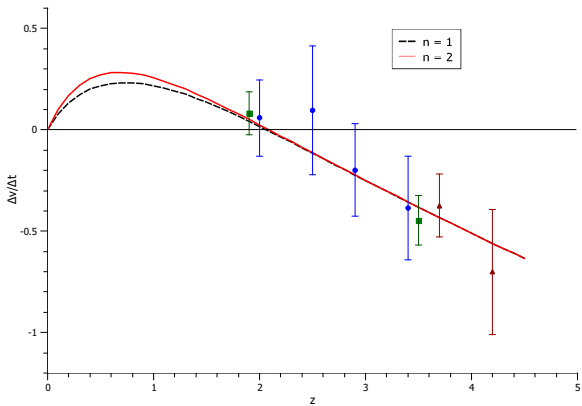}
%}
\caption{Variation of velocity drift (cm/sec/yr) with redshift $z$ for $f(R)$
gravity model. }\label{fig:fR}
\end{figure}

The Fig. \ref{fig:fR} displays the variation of velocity drift for
$f(R)$ model. There is very small difference between the variation
of $H(z)$ with z for n =1 and n = 2 at the redshift $ z \le 1.8$
\cite{dev}. Hence for redshift $ z \ge 1.8$, the variation of
velocity drift becomes independent of the value of $n$.

\section{Results and Discussions}

Till now, large number of theoretical models have been proposed to
explain the observed accelerated expansion of the universe. In order
to get insight of the mechanism behind this acceleration, we require
complimentary observational tools. The observational tests either
belong to {\em distance based} methods such as SNe Ia luminosity
distances, angular size of compact radio sources, BAO, CMBR,
gravitational lensing  etc. or {\em time based} methods like
Absolute age method, Lookback time method and  Differential age
method. Every observational tests  requires some priors, assumptions
and is subjected to systematic errors.

There is a need to develop tools which are simple and have a
controlled systematics. The cosmological redshift drift (CRD) method
is a test in this direction. Corasaniti et al. were the first to
analyse dark energy models like $\Lambda$CDM, Chaplygin gas and dark
energy- dark matter interaction model using the CRD test~\cite{cor}.
They conclude that the CRD test puts stringent constraints on
non-standard dark energy models. Later on, Balbi and Quercellini
also investigated various standard and non-standard dark energy
models~\cite{ba}. They found  the worst bound for the Cardassian
model.  Further Zhang et al. also studied Holographic dark energy
model with CRD test~\cite{zh} and obtain a very tight bound on the
$\Omega_m$.  The above stated work used the data generated by MC
simulations, given by Pasquini et al. (2006)~\cite{pa}.

Uzan, Bernardeau and Mellier discuss the possibility in which the
large scale structure may effect the measurements of  time drift of
the cosmological redshift~\cite{u}. Since the CRD test is based on
the assumption that the universe is homogeneous and isotropic,
therefore this test has been used to check the homogeneity of the
universe~\cite{u1}. Quartin and Amendola propose that CRD test can
be used to distinguish between Void models and conventional dark
energy scenarios~\cite{q}. Recently one possible source of noise in
the CRD test has been studied by Killedar and  Lewis which deal with
the transverse motion of the Ly$\alpha$ absorbers \cite{kill}.

Recently Liske et al. (2008) studied in detail the impact of next
generation ELT on observing the very small redshift drift
signal~\cite{lis}. Using the extensive MC simulations, three sets of
data points are generated with  different strategies which include
the measurement of the precise value of velocity shift. In  our
work, We used the recent data generated by Liske at al.(2008) to
constrain the late time acceleration models of the universe. We
study the models which belong to both the Einstein gravity scenario
($\Lambda$CDM, XCDM, scalar field potentials) and the modified
gravity ($f(R)$).

Results are summarized as follows:

\begin{enumerate}

\item In $\Lambda$CDM model $(w_x = -1)$, the $\chi^2$ minimum lies
at $\Omega_{m0}= 0.3$. Considering $h$ to be a nuisance parameter, we
marginalize over $h$ to obtain the probability distribution function
defined as:

$$
L(p)\, = \, \int{e^{-\chi^2(h,p)/2}P(h)dh}
$$

Here P(h) is the prior probability function for $h$ which is assumed
to be Gaussian. We find that the best fit value of model  parameter is
independent of the choice of prior.

\item We performed the $\chi^2$ statistics on flat XCDM dark
energy model as shown in Fig. \ref{contour}. The best fit values
lies at  $\Omega_{m} = 0.3$ and  $ w_x = - 1$. At $1 \sigma $-level
the constraints are,
\[ \Omega_m \, = \, 0.30^{+0.03}_{-0.04} \hskip 2 cm \omega_x \le
-0.62
\]
Here again we marginalize over $h$ to obtain the best fit value of
model  parameters $(\Omega_m, w_x)$. The best fit values are
independent of the choice of prior. The redshift drift test give
very tight constraint on the $\Omega_m$ and weak bound on the
equation of state, $\omega_x$. This is expected since the amplitude
and slope of the cosmic velocity shift w.r.t the redshift is very
sensitive to the $\Omega_m$ and shows weak dependence on the
equation of state. The other important feature of this analysis is
that it is  complementary to other probes such as CMB, BAO and weak
lensing.

\begin{figure}[ht]
\centering
\includegraphics[width=6.5cm]{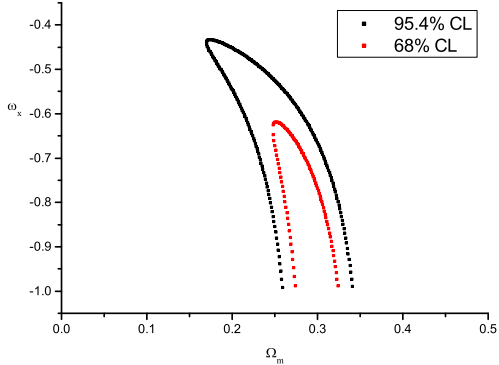}
\caption{Flat dark energy model with constant $\omega_x$: Contours
in $\Omega_m$ and $\omega_x$ plane for CRD test. The best fit value
lies at $ \Omega_m= 0.3$ and  $\omega_x = -1$. The inner and outer
curves are at $1 \sigma$ and $2 \sigma$ respectively.
}\label{contour}
\end{figure}

\item We analyze the $f(R)$ gravity model proposed by
Starobinsky  with the CRD test. The variation of velocity drift for
different values of $n$ with  redshift $z$ is shown in Fig.
\ref{fig:fR}. The expansion history of this model match exactly with
the $\Lambda$CDM model for $ n = 2$ and $\lambda = 2$. In this model
the Hubble parameter, $H(z)$, becomes independent of the parameter
$n$ after the redshift $z = 1.8$ and expansion history traces
exactly the $\Lambda$CDM model behavior at high redshift~\cite{odintsov,dev}.
Since the simulated data points have $z \geq 1.9$, the $\chi^2$ for
this model will be the same as for the $\Lambda$CDM model. Here
$\Lambda$CDM model means standard cosmological model with $\omega_x
= -1$, $\Omega_{m0}= 0.3$ and $\Omega_{\Lambda 0}= 0.7$ . The
effective equation of state for this model appproaches $ \omega = -1$
at the present epoch [see Fig.8 of the ref.\cite{dev}].
 In order
to constrain this model better we need redshift drift data in the
redshift range $ z \le 1.9$ .

\item In Fig. \ref{chi_exp}, we plot the variation of $\chi^2$ with the
parameter $\beta$ for the exponential potential. The best fit value
of $ \beta = m_{pl}^2/(8\pi f^2)=3.2$ corresponds to $ \Omega_{m0} =
0.26$. The $\chi^2$ per degree of freedom is 0.67.  At $3 \sigma$
limit, the allowed range of $ \beta = 3.2^{-0.45}_ {+0.55}$ gives $
\Omega_m = 0.26^ {-0.05}_{+0.06}$. Although the model predicts the
observed value of $\Omega_m$ but still this model is not fully in
concordance with the observations; as  the allowed range of $\beta$
at $3 \sigma$ level shows that the universe mostly remains in a
decelerating phase (see Fig. \ref{fig:exp-pot}).

\begin{figure}[ht]
\centering
\includegraphics[width=5.2 cm,angle=-90]{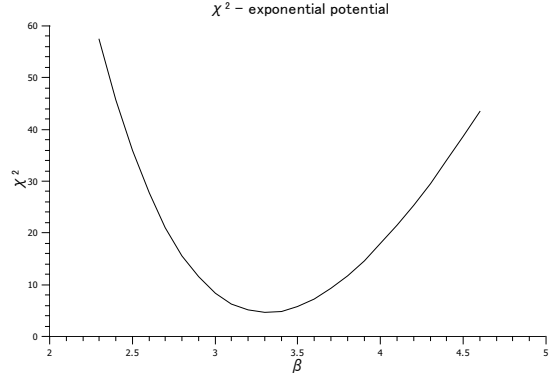}
\caption{Variation of $\chi^2$ with $\beta$}\label{chi_exp}
\end{figure}

\item  In the inverse power law potential model, for $\gamma \rightarrow
0$, the energy-momentum tensor approaches towards cosmological
constant. The best fit values for this model are  $\gamma =0.025$
and $K = 0.118 $. In Fig. \ref{Contour_Power_law} the contour is
plotted in $ \gamma - K$ plane. We would like to note here that
computational efforts to generate the data points in the
neighborhood of $\gamma = 0$ increases. The important feature of
this contour is that it is complementary to the contour obtain by
using the Dark Energy Task Force (DETF) simulated data sets for
future experiments. This includes SNe Ia, BAO, weak Lensing and CMB
(PLANCK) observations \cite{y}.

In the Fig. \ref{eqn_power_law}  variation of instantaneous equation of state
is plotted w.r.t. redshift for the best fit values of the model
parameters. There is almost no variation in the equation of state
with redshift. It always stay close to $ \omega = -1$  for the
redshift range studied here.

\begin{figure}[ht]
 \centering
\includegraphics[width=6.0cm,angle=0]{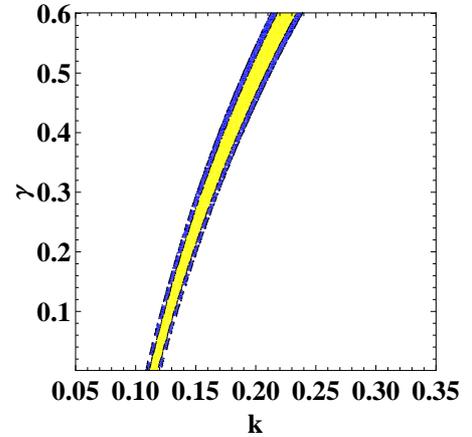}
\caption{Contour in  $\gamma - K$ plane for inverse  power law potential.  The inner and the outer contours are  drawn at the  1$\sigma$ and 2$\sigma$ level respectively.}\label{Contour_Power_law}
\end{figure}

\begin{figure}[ht]
 \centering
\includegraphics[width=6.0cm,angle=0]{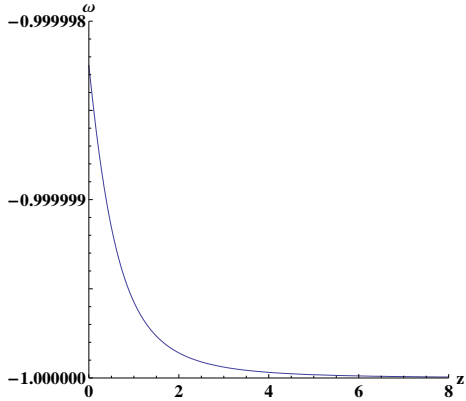}
\caption{Variation of $\omega$ w.r.t. redshift for inverse  power law
potential.}\label{eqn_power_law}
\end{figure}

\item For the PNGB model, the contour is plotted in $\alpha - m$ plane
as shown in the Fig. \ref{Contour_PNGB}. The $\chi^2$ minimum lies
at $\alpha =0.21 $ for $ m = 0.56$. At 2$\sigma$ level, $\alpha >
0.15$ is allowed. This result is completely  in concordance with the
other cosmological observations (eg. SNe Ia, lensing, cluster
observations) \cite{waga,ng,wa}. Again this contour is complementary
to the contour obtain by using Markov  Chain Monte Carlo analysis
(MCMC) to generate the future data set. This includes future SNe Ia,
BAO, weak Lensing and cosmic microwave background observations
\cite{aa}.

In the Fig. \ref{eqn_pngb_law}
instantaneous equation of state  is plotted for the
best fit values. In contrast to power law potential, the PNGB show
large variation in equation of state for the redshift $ z < 2$.

\begin{figure}[ht]
\centering
\includegraphics[width=5.30 cm,angle=0]{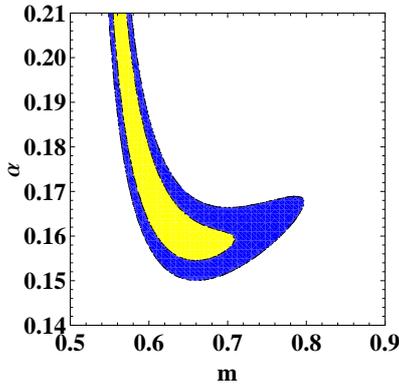}
\caption{Contour in  $\alpha - m$ plane for PNGB potential. The
inner and the outer contours are  drawn at the  1$\sigma$ and
2$\sigma$ level respectively.}\label{Contour_PNGB}
\end{figure}

\begin{figure}[ht]
 \centering
\includegraphics[width=5.30cm,angle=0]{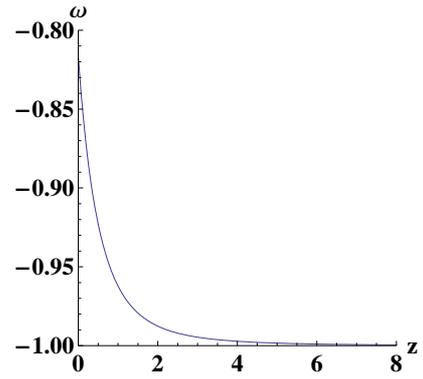}
\caption{Variation of  $\omega $ w.r.t. redshift
for for PNGB potential.}\label{eqn_pngb_law}
\end{figure}

It is clear that CRD test is a very simple, straightforward and
powerful tool which probes the expansion history of the universe
directly. In future the precision data especially in redshift range
of $z < 2$ shall add to the predictive power of CRD test.

\end{enumerate}

\section*{Acknowledgment}

We are thankful to Joe Liske and Abha Dev for  useful
discussions at the various stages of this work. Authors also
acknowledges the hospitality provided by the IUCAA, Pune where part
of the work is done. One of the author (DJ) also thank A. Mukherjee
and S. Mahajan for providing the facilities to carry out the
research. Finally authors acknowledge the financial support provided
by Department of Science and Technology, Govt. of India under project No.
SR/S2/HEP-002/2008.

\bibliographystyle{elsarticle-num}

\end{document}